\documentclass[notitlepage,10pt,prfluids,aps,superscriptaddress]{revtex4-2}
\usepackage{easymath}
\usepackage{graphicx}
\usepackage{epstopdf}
\usepackage{tabularx}
\usepackage{amsmath}
\usepackage{url}
\usepackage{enumitem}
\usepackage[colorlinks]{hyperref}
\hypersetup{
    citecolor = {magenta}
}

\makeatletter
\newcommand\xleftrightarrow[2][]{%
  \ext@arrow 9999{\longleftrightarrowfill@}{#1}{#2}}
\newcommand\longleftrightarrowfill@{%
  \arrowfill@\leftarrow\relbar\rightarrow}
\makeatother

\begin{document}

\title{Vortices of Electro-osmotic Flow in Heterogeneous Porous Media}

\author{Mohammad Mirzadeh}
\affiliation{Department of Chemical Engineering, Massachusetts Institute of Technology, MA 02139.}

\author{Tingtao Zhou}
\affiliation{Department of Physics, Massachusetts Institute of Technology, MA 02139.}

\author{Mohammad Amin Amooie}
\affiliation{Department of Chemical Engineering, Massachusetts Institute of Technology, MA 02139.}

\author{Dimitrios Fraggedakis}
\affiliation{Department of Chemical Engineering, Massachusetts Institute of Technology, MA 02139.}

\author{Todd R.\ Ferguson}
\affiliation{Aramco Americas Company: Aramco Research Center-Boston, 400 Technology Square, Cambridge, MA 02139.}

\author{Martin Z.\ Bazant}
\email[Corresponding author: ]{bazant@mit.edu}
\affiliation{Department of Chemical Engineering, Massachusetts Institute of Technology, MA 02139.}
\affiliation{Department of Mathematics, Massachusetts Institute of Technology, MA 02139.}

\date{\today}

\begin{abstract}
Traditional models of electrokinetic transport in porous media are based on homogenized material properties, which neglect any macroscopic effects of microscopic fluctuations.  This perspective is taken not only for convenience, but also motivated by the expectation of irrotational electro-osmotic flow, proportional to the electric field, for uniformly charged surfaces (or constant zeta potential) in the limit of thin double layers.  Here, we show that the inherent heterogeneity of porous media generally leads to macroscopic vortex patterns, which have important implications for convective transport and mixing. These vortical flows originate due to competition between pressure-driven and electro-osmotic flows, and their size are characterized by the correlation length of heterogeneity in permeability or surface charge.   The appearance of vortices is controlled by a single dimensionless control parameter, defined as the ratio of a typical electro-osmotic velocity to the total mean velocity.
\end{abstract}

\maketitle

\section{Introduction}
Flows in porous media are everywhere around us \cite{bear2013dynamics}. From the small scale of nutrient and heat transport in biological tissues \cite{khaled2003role} to the geological scale of subsurface flows \cite{sahimi2011flow}, involved transport processes share common principles. Many transport processes in porous media can be recast in terms of driving forces and corresponding fluxes. This idea is at the heart of non-equilibrium thermodynamics, in which fluxes and driving forces are coupled through Onsager relations that enforce symmetry of the linear response matrix \cite{de2013non}. The nonlinear coupling between various driving forces can result in rich, and sometimes unexpected, behavior.  Indeed, recent works indicate electrokinetic phenomena in porous media could be exploited to control viscous fingering  \cite{mirzadeh2017electrokinetic,gao2019active}, suppress instabilities in the growth of nano-wires \cite{han2014over,shin2014race} and porous electrodeposits \cite{han2016dendrite}, and drive over-limiting current \cite{dydek2011overlimiting,yaroshchuk2011coupled} and deionization shock waves \cite{zangle2010theory,mani2011deionization,nam2015experimental}, which enable water purification by shock electrodialysis \cite{deng2013overlimiting,deng2015water,schlumpberger2015scalable,conforti2020continuous,alkhadra2019continuous,alkhadra2020small}.

A unique aspect of transport in porous media is the role of randomness in the pore geometry and network topology \cite{sahimi2011flow}. A familiar example is flow channeling in a heterogeneous medium. When subjected to pressure gradient, fluids preferentially flow along the path of least resistance and avoid regions of low permeability, leading to the formation of ``flow channels''. Similarly, in two-phase flows, randomness leads to the formation of two well-known phenomena, \ie the viscous fingering and capillary fingering \cite{homsy1987viscous, lenormand1988numerical}. These instabilities can increase mixing and are often deemed undesired, \eg in secondary oil recovery \cite{amooie2017hydrothermodynamic}. Quantifying mixing in porous media is also critical to understanding reactive transport \cite{steefel2005reactive, dentz2011mixing}, as well as dissolution trapping in pore fluid following geologic CO$_2$ sequestration  \cite{amooie2018solutal, mahmoodpour2019convective}.

``Passive'' control of fluid flow is possible via careful geometrical manipulation, \eg in patterned micro-fluidic devices to enhance mixing \cite{stroock2002patterning, stroock2002chaotic}, or in Hele-Shaw cells \cite{al2012control} and porous media \cite{rabbani2018suppressing} to suppress interfacial instabilities. In passive control, the extent of flow manipulation is limited and difficult to adjust externally or dynamically. Conversely, ``active'' control may be possible by exploiting the coupling between competing driving forces. One possibility is using electric fields to manipulate fluid flow via electrokinetic phenomena \cite{hunter2013zeta}. This idea was recently shown to enable the active control of viscous fingering \cite{mirzadeh2017electrokinetic,gao2019active} by modifying the effective hydraulic resistance experienced by the fluids.

Traditional models of electro-osmosis in porous media and micro-fluidic devices assume irrotational flow with strongly screened hydrodynamic interactions \cite{long2001note}, based on formal homogenization \cite{schmuck2015homogenization} or the mathematical limit of thin double layers and uniform zeta potential, in which the fluid velocity is proportional to the (irrotational) electric field \cite{cummings1999irrotationality,cummings2000conditions}.  Indeed, the assumption of uniform electro-osmotic flow driven by a uniform electric field underlies models of various industrial processes, such as electrokinetic soil remediation and ionic separations \cite{shapiro1993removal,probstein1993removal,acar1993principles,alshawabkeh1996electrokinetic,virkutyte2002electrokinetic}.  Even in situations of stochastic electrotransport with fluctuating electric fields \cite{kim2015stochastic}, used to accelerate chemical transport for the imaging of biological tissues and organs \cite{murray2015simple,ku2016multiplexed,mano2018whole} , the instantaneous velocity field is assumed to be spatially uniform. The limit of thin double layers is also invoked to justify the approximate independence of electrophoretic mobility on particle shape \cite{morrison1970electrophoresis,oss1975influence,teubner1982motion,anderson1989colloid}, which makes particle separation challenging, unless symmetry is broken by nonuniform \cite{long1998symmetry,long1996electrophoretic} or induced \cite{bazant2004induced,squires2006breaking} surface charge. Inhomogeneity in surface charge or shape can further be utilized for steady pumping \cite{ajdari1995electro, ajdari2000pumping} or patterning flow fields in micro-fluidic devices \cite{stroock2000patterning, long1999electroosmotic}.  Recently, similar ideas have been proposed for pattering flow fields in Hele-Shaw cells and micro-fluidic devices using gate electrodes \cite{boyko2015flow, paratore2019dynamic, paratore2019electroosmotic}, although such a strategy would be difficult to achieve in a porous medium.

Here, we demonstrate that a heterogeneous permeability field has a similar effect on pattering the flow field and creating vortical flows, which are beneficial for enhanced mixing. This phenomenon occurs due to strong internal pressure generated by the electro-osmotic flow. Although the concept of electro-osmotic flow reversal is understood for individual pores \cite{hunter2013zeta}, here we demonstrate similar patterns at the much larger macroscopic scale. Through detailed analyses, we show that the size of vortical structures directly scale with the length scale of heterogeneity in the domain. More importantly, we find that the structure of flow field can be described in terms of the ``electro-osmotic coupling coefficient'', a nondimensional parameter measuring the relative strength of electro-osmotic velocity to total mean velocity and which is tunable experimentally. Finally, recent work on coupled nonlinear electrokinetics in random pore network also demonstrate similar vortical flow structures and their interactions with deionization shock waves \cite{alizadeh2019impact}. In this article, we neglect nonlinear effects and instead provide a clear physical explanation for the formation of vortical structures.

\section{Physical Picture} \label{sec:physical-picture}
\subsection{Electrokinetic Phenomena and Electro-osmotic Pumping}
\begin{figure}
    \centering
    \includegraphics[width=0.75\columnwidth]{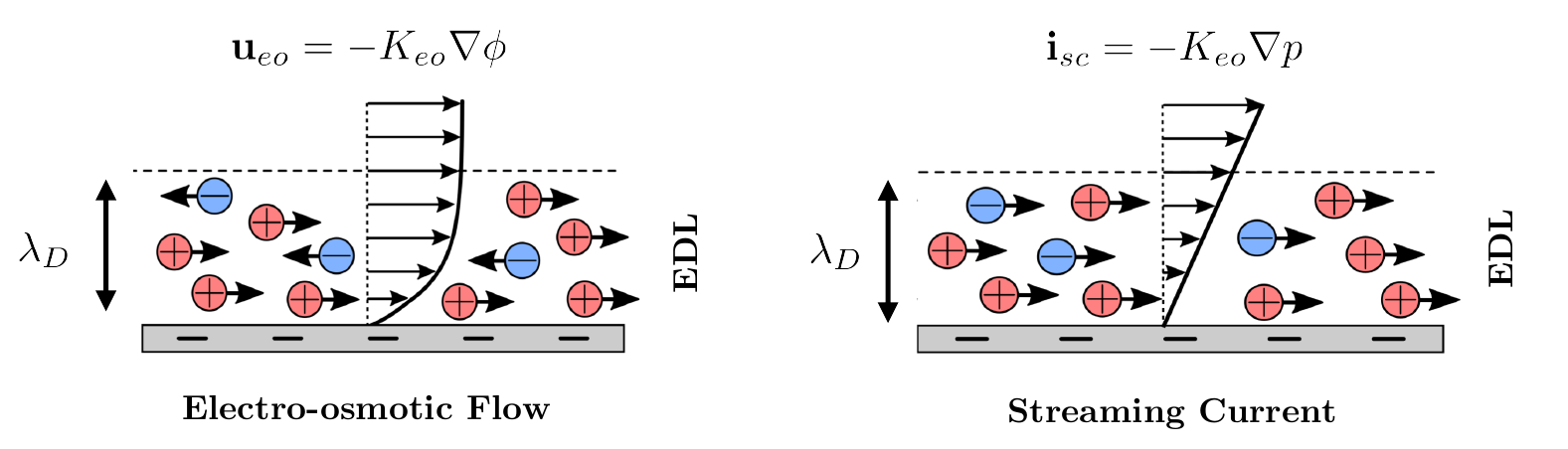}
    \caption{Many surfaces are charged in aqueous solution. When ions are present in the solution, a thin cloud of mostly counter-ions accumulate near the surface to screen the surface charge and form the Electric Double Layer (EDL). \textbf{Left:} An electric field drives electro-osmotic flow due to electrostatic force on the ions in the EDL. \textbf{Right:} Conversely, streaming current occurs due to advection of charges in the EDL by the pressure-driven flow.}
    \label{fig:schematic-edl}
\end{figure}
Many surfaces are charged in contact with aqueous solutions due to the dissociation of surface groups, \eg Silanol in glass or silicate minerals. When ions are present, the surface charge is screened by a diffuse cloud of equal and opposite charges, mostly counter-ions, which form the so-called the Electric Double Layer (EDL). The EDL is often characterized by its thickness, the Debye length ($\lambda_D$), which, for a binary electrolyte, is:
\begin{equation}
    \lambda_D = \sqrt{\frac{\varepsilon k_\mathrm{B}T}{2 c_0 z^2 e^2}}.
    \label{eq:debye-length}
\end{equation}
Here, $\varepsilon$ is the permittivity of the electrolyte, $k_\mathrm{B}$ is the Boltzmann constant, $T$ is the absolute temperature, $c_0$ is the salt concentration, $z$ is the ion valence, and $e$ is the elementary charge. For $c_0 = 1 \,\mathrm{mM}$ solution of mono-valent ions, $z=1$, at room temperature, the Debye length is very small, $\lambda_D \approx 10\,\mathrm{nm}$. Despite being very thin, the interactions between the ions in the EDL and solvent molecules, typically water, lead to a myriad of transport processes that are collectively termed ``electrokinetic phenomena'' \cite{hunter2013zeta}. In particular, the application of an external electric field parallel to the surface exerts electrostatic forces on the ions and drives ``electro-osmotic flow'' (see figure \ref{fig:schematic-edl}). The electro-osmotic velocity reaches a constant value away from the surface, which, for small driving forces, scales linearly with the electric field $\mathbf{E} = -\nabla \phi$:
\begin{equation}
    \mathbf{u}_{eo} = K_{eo} \mathbf{E} = -K_{eo} \nabla \phi,
    \label{eq:ueo}
\end{equation}
where $\phi$ is the electrical potential and $K_{eo}$ is the electro-osmotic mobility. Similarly, advection of ions in the EDL due to pressure-driven flow generates ``streaming current'', with an average density $\mathbf{i}_{sc} = -K_{sc} \nabla p$. When the driving forces are small, the Onsager symmetry, a postulate of linear irreversible thermodynamics, requires that electrokinetic phenomena are symmetric with respect to driving forces, \ie $K_{sc} = K_{eo}$. When the EDL is thin compared to a geometrical length-scale, \eg the pore size in a porous medium or the gap size in a Hele-Shaw cell, the electro-osmotic mobility is given by the Helmholtz-Smoluchowski relation:
\begin{equation}
    K_{eo} = \frac{-\varepsilon \zeta}{\mu},
    \label{eq:keo}
\end{equation}
where $\mu$ is the electrolyte viscosity and $\zeta$ is the potential difference across the EDL, which depends on several variables including surface charge density, electrolyte concentration, and pH. For many surfaces, the $\zeta$-potential varies in the range of $-100 < \zeta < 50 \,\mathrm{mV}$ at room temperature \cite{hunter2013zeta, kirby2004zeta}.

When both pressure gradients and electric fields are present, the total average velocity is the sum of hydraulic and electro-osmotic components:
\begin{equation}
    \mathbf{u} = \mathbf{u}_h + \mathbf{u}_{eo} = -K_h \nabla p  - K_{eo} \nabla \phi,
    \label{eq:utotal-scaling}
\end{equation}
where the hydraulic conductivity is given by the Darcy relation, $K_h = k/\mu$, and $k$ is the Darcy permeability. Equation \eqref{eq:utotal-scaling} could be understood as the macroscopic velocity in a porous medium, the area-averaged velocity in a cylindrical pore, or the depth-averaged velocity in a Hele-Shaw cell. When electro-osmotic flows are driven into tight pores, strong adverse pressure gradients are generated if the pore cannot sustain the flow rate. This is a consequence of different scaling of hydraulic and electro-osmotic velocities with the characteristic length-scale $h \sim \sqrt{k}$. While the electro-osmotic velocity is independent of geometrical length-scales, the hydraulic velocity scales quadratically, $U_h \sim h^2$. From equations \eqref{eq:keo} and \eqref{eq:utotal-scaling}, the maximum pressure gradient occurs when the total velocity is zero, \eg in a dead-end pore:
\begin{equation}
    \Delta p_{eo} \sim \frac{\varepsilon \zeta}{h^2} \Delta \phi.
    \label{eq:peo}
\end{equation}

\begin{figure}[t]
    \centering
    \includegraphics[width=0.8\columnwidth]{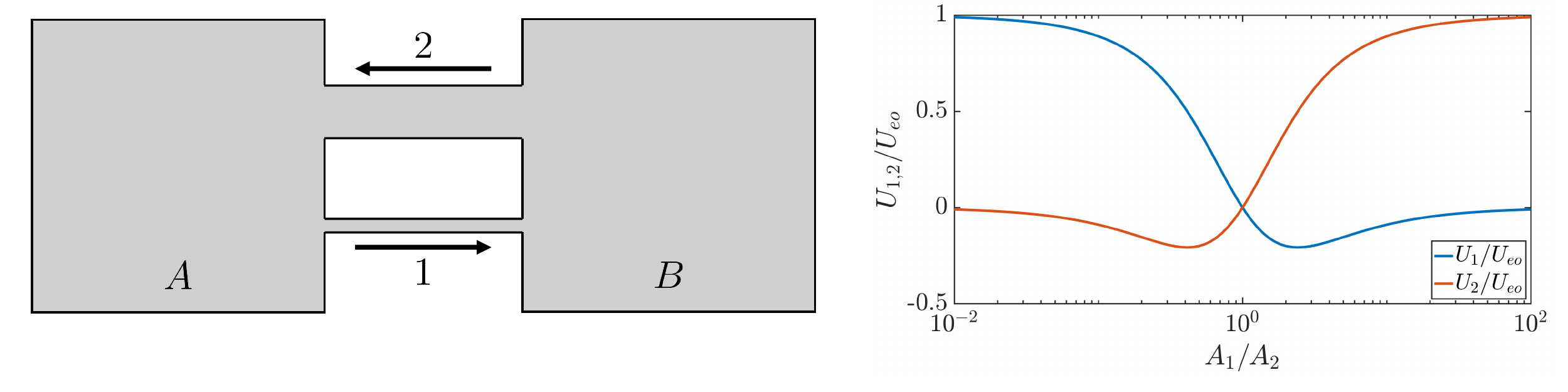}
    \caption{Flow reversal occurs when pressure-driven and electro-osmotic flows compete. \textbf{Left:} Schematic of two reservoirs `A' and `B', connected via two parallel ``pores'' of different sizes $A_1$ and $A_2$. \textbf{Right:} When an electric field is applied from reservoir `A' to `B', electro-osmotic flow creates an adverse pressure gradient, which drives backward pressure-driven flow. Because the hydraulic and electro-osmotic conductivities scale differently with the pore size, the flow in the smaller pore is always in the opposite direction of the larger pore, leading to a net circulation.}
    \label{fig:schematic-circulation}
\end{figure}

Equation \eqref{eq:peo} indicates that electro-osmotic flow can ``pump'' the fluid in the opposite direction of pressure gradient. It is well known that electro-osmotic pumping leads to flow reversal in a dead-end pore or near a bottle-neck \cite{park2006eddies}, which is a possible mechanism for sustaining over-limiting currents \cite{dydek2011overlimiting, yaroshchuk2011coupled}. Adverse pressure gradient and internal flow re-circulation also limit the performance of electro-osmotic pumping, which must be circumvented to achieve fast pumping \cite{urbanski2006fast, bazant2006theoretical, huang2010ultrafast}. Consider a hypothetical situation where two reservoirs, `A' and `B', are connected via a small and large pore placed in parallel (see figure \ref{fig:schematic-circulation}). If an electric field is applied from reservoir `A' to `B', the electro-osmotic flow creates an adverse pressure difference:
\begin{equation}
    \Delta p_{eo} = p_B - p_A = \frac{R_1 R_2}{R_1 + R_2}U_{eo} \left(A_1 + A_2\right),
\end{equation}
where $A_{1,2}$ and $R_{1,2}$ are the cross-sectional area and hydraulic resistance of the pores, respectively. This pressure difference drives hydraulic flow in both pores that are in the opposite direction of electro-osmotic flow, leading to the total area-averaged velocities given by:
\begin{align}
    U_1 &= U_{eo} - \frac{\Delta p_{eo}}{R_1 A_1} = U_{eo}\left(1 - \frac{R_2}{R_1 + R_2} \frac{A_1 + A_2}{A_1}\right), \label{eq:U1} \\
    U_2 &= U_{eo} - \frac{\Delta p_{eo}}{R_2 A_2} = U_{eo}\left(1 - \frac{R_1}{R_1 + R_2} \frac{A_1 + A_2}{A_2}\right). \label{eq:U2}
\end{align}

From Darcy's law, $R_2/R_1 = (A_1/A_2)^2$, and therefore average velocities may be expressed in terms of relative pore size, $A_1/A_2$. Figure \ref{fig:schematic-circulation} illustrates that the average velocity in the larger pore is always in the opposite direction of the smaller pore. This is easy to understand in the limit when $A_1 \ll A_2$. In this case, the electro-osmotic flow primarily reverses through the pore with least hydraulic resistance (larger area). The simple analysis presented here is the basis of flow reversal and circulation in a random porous medium as will be shown in the following section. A porous medium with a heterogeneous permeability field may be idealized using a collection of different-sized pores that are connected in series and parallel. While series connection can lead to flow reversal at the pore scale, it is the size variation between parallel pores that lead to flow reversal at the macroscopic level.

It is well known that pore network connectivity impacts fluid flow \cite{chen1985pore, lenormand1988numerical, holtzman2016effects}, ion transport \cite{mirzadeh2014enhanced}, and electrokinetics \cite{alizadeh2017multiscale, alizadeh2019impact} in porous media. Recently, the concept of ``accessivity'' \cite{gu2019microscopic} was introduced to characterize the role of pore network connectivity in explaining the origins of capillary hysteresis in porous media. This concept has also been used  to quantify the impact of network heterogeneity on sustaining over-limiting currents through ion-selective membranes \cite{alizadeh2019impact}. Therefore, it should not be surprising that pore network connectivity also impacts vortex formation. As we shall see, the circulation region in a heterogeneous porous medium roughly scales with the size of heterogeneity, \ie the length-scale over which the permeability field changes in the domain. However, before turning to porous media, we will first consider a simpler problem in a Hele-Shaw cell with nonuniform gap thickness which allows for analytical solution.

\subsection{nonuniform Hele-Shaw Cell}
\begin{figure}
    \centering
    \includegraphics[width=0.9\columnwidth]{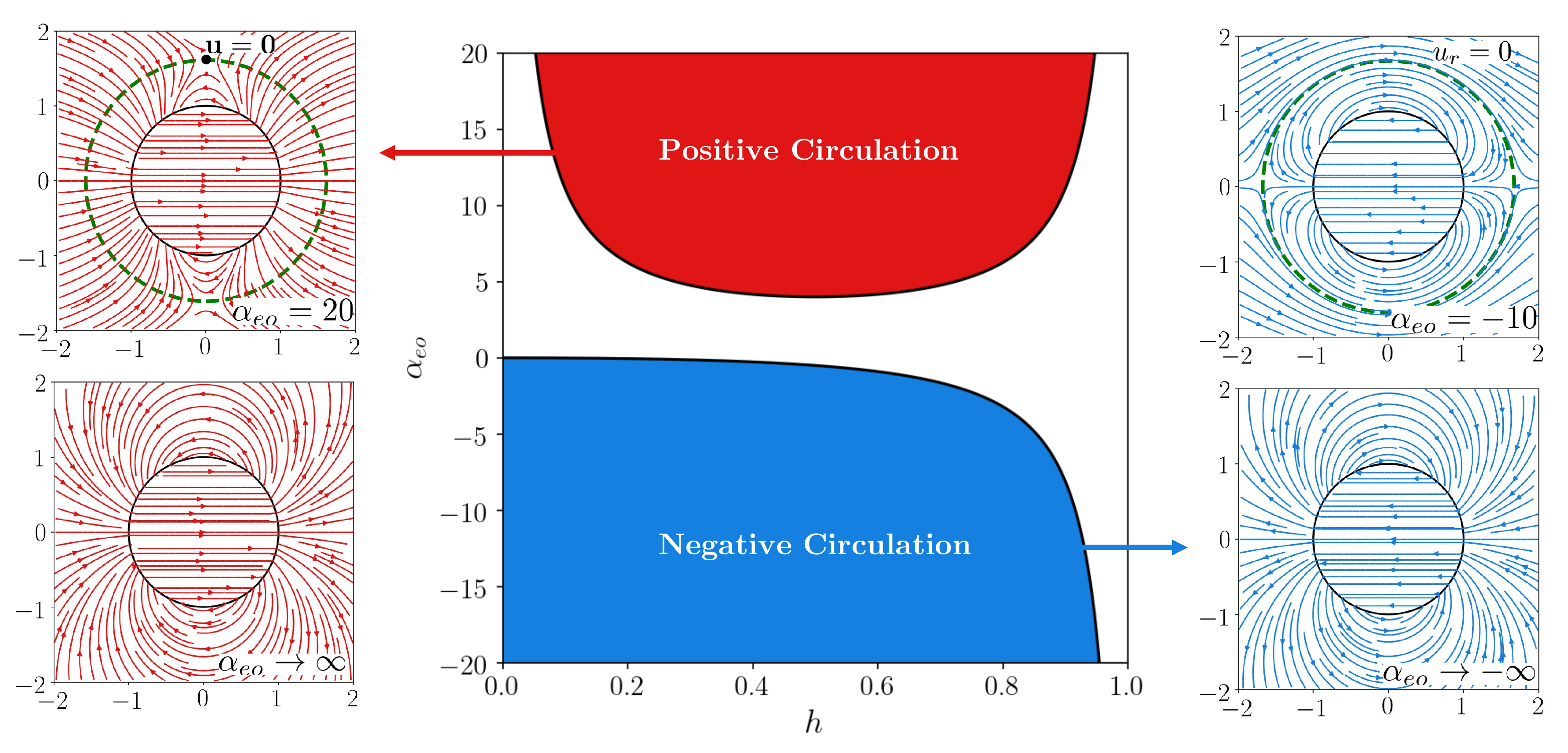}
    \caption{Flow characterization in a Hele-Shaw cell of variable gap. \textbf{Center:} The state of flow field is fully determined by two parameters: the gap ratio, $h$, and the electro-osmotic coupling coefficient, $\alpha_{eo}$. For any gap ratio, circulation is possible if the electro-osmotic velocity is sufficiently strong. The direction of electro-osmotic flow, denoted by the sign of $\alpha_{eo}$, dictates the direction of the dipolar flow field and the sign of circulation. \textbf{Left, Right:} Representative examples of the flow field for the gap ratio of $h=0.1$. The green dashed line represents the extent of circulation region given via equations \eqref{eq:HS-alpha-neg}, and \eqref{eq:HS-alpha-pos} which grows in an unbounded fashion as $|\alpha_{eo}|\rightarrow\infty$.}
    \label{fig:hele-shaw}
\end{figure}

Flows in Hele-Shaw cells have been traditionally studied as a two-dimensional idealization of more complex flows in porous media. Moreover, flows in Hele-Shaw cells are considerably easier to visualize experimentally. Recently, several works have shown that by controlling the zeta potential through gate electrodes, it is possible to control flow patterns in a uniform Hele-Shaw cell \cite{boyko2015flow, paratore2019dynamic, paratore2019electroosmotic}. These experiments clearly illustrate the formation of circulation regions around gate electrodes. Here, we consider a Hele-Shaw cell with a disk-like region of narrower gap thickness, but constant zeta potential (see figure \ref{fig:hele-shaw}). As we will see, the depth-averaged velocity for this problem also exhibits similar circulating flow pattern around the disk region.

We assume a Hele-Shaw cell with a variable gap thickness $H(\mathbf{x}) = H_0 - \Delta H \chi(\mathbf{x})$, where $\chi = 1$ inside the disk of radius $a$ and $\chi = 0$ outside. The Hele-Shaw cell is subjected to uniform fluid flow and electric current far away from the disk. The presence of the nonuniformity in the gap thickness affects the hydraulic and electrical resistance near the disk and perturbs the pressure and electric potential fields. To obtain the depth-averaged velocity, we must first solve for the electric potential which satisfies the depth-averaged Ohm's law:
\begin{equation}
    \divgp{H(\mathbf{x}) \sigma \nabla \phi} = 0.
    \label{eq:HS-potential}
\end{equation}
Here, $\sigma$ is the electrolyte conductivity which is assumed to be uniform. Notice that we have ignored the contribution from streaming current in equation \eqref{eq:HS-potential}. This assumption is justified for Hele-Shaw cells in which the gap thickness is considerably larger than the Debye length ($H \gg \lambda_D$). We will further comment on this assumption in the next section where we define a nondimensional parameter to quantify the strength of streaming current. The solution to equation \eqref{eq:HS-potential} in polar coordinates is:
\begin{align}
    \phi_1 &= -E_\infty r\cos\theta \left(\frac{2}{1 + h}\right),
    \label{eq:HS-phi_inside} \\
    \phi_2 &= -E_\infty r\cos\theta \left(1 + \frac{a^2}{r^2} \frac{1 - h}{1 + h}\right),
    \label{eq:HS-phi-outside}
\end{align}
where $E_\infty$ is the far-field electric field, $h = H_1/H_2$ is the gap nonuniformity parameter, and $\phi_1$ and $\phi_2$ are the solution inside and outside the disk, respectively. Using this solution, the electro-osmotic velocity is given by:
\begin{align}
    \mathbf{u}_{eo}^1 &= \frac{2  U_{eo}}{1+h} \left(\cos\theta \,\mathbf{\hat{e}}_r -\sin\theta \,\mathbf{\hat{e}_\theta} \right),
    \label{eq:HS-ueo-inside} \\
    \mathbf{u}_{eo}^2 &= U_{eo} \cos\theta \left(1 - \frac{a^2}{r^2}\frac{1-h}{1+h}\right) \mathbf{\hat{e}}_r - U_{eo} \sin\theta \left(1 + \frac{a^2}{r^2}\frac{1-h}{1+h}\right) \mathbf{\hat{e}_\theta},
    \label{eq:HS-ueo-outside}
\end{align}
where $U_{eo} = K_{eo} E_\infty$ is the far-field electro-osmotic velocity.

For a Hele-Shaw cell, equation \eqref{eq:utotal-scaling} gives the total depth-averaged velocity with $k(\mathbf{x}) = H(x)^2/12\mu$. Mass conservation then requires that:
\begin{equation}
    \divgp{\frac{H(\mathbf{x})^3}{12 \mu} \nabla p} + \divgp{H(\mathbf{x}) K_{eo} \nabla \phi}= 0,
\end{equation}
which combined with equation \eqref{eq:HS-potential} further simplifies to:
\begin{equation}
    \divgp{\frac{H(\mathbf{x})^3}{12 \mu} \nabla p} = 0.
    \label{eq:HS-pressure}
\end{equation}
Note that the apparent decoupling between the pressure and potential fields in equation \eqref{eq:HS-pressure} is simply a result of assuming uniform electro-osmotic mobility and electric conductivity. As we will discuss in the next section, this assumption may not hold in general. Nevertheless, the coupling between the two fields is still enforced through the boundary conditions. The solution to equation \eqref{eq:HS-pressure}, subject to uniform far-field velocity, is given by:
\begin{align}
    p_1 &= -\frac{12 \mu U_h}{H_0^2} r\cos\theta \left(\frac{2}{1 + h^3}\right),
    \label{eq:HS-p-inside} \\
    p_2 &= -\frac{12 \mu U_h}{H_0^2} r\cos\theta \left(1 + \frac{a^2}{r^2} \frac{1 - h^3}{1 + h^3}\right),
    \label{eq:HS-p-outside}
\end{align}
where $U_h = U - U_{eo}$ is the hydraulic part of the uniform far-field velocity, $U$. From equations \eqref{eq:HS-p-inside} and \eqref{eq:HS-p-outside}, the hydraulic velocity is found as:
\begin{align}
    \mathbf{u}_{h}^1 &= \frac{2 h^2 U_{h}}{1+h^3} \left(\cos\theta \,\mathbf{\hat{e}}_r - \sin\theta \,\mathbf{\hat{e}_\theta} \right),
    \label{eq:HS-uh-inside} \\
    \mathbf{u}_{h}^2 &= U_{h} \cos\theta \left(1 - \frac{a^2}{r^2}\frac{1-h^3}{1+h^3}\right) \,\mathbf{\hat{e}}_r - U_{h} \sin\theta \left(1 + \frac{a^2}{r^2}\frac{1-h^3}{1+h^3}\right) \,\mathbf{\hat{e}_\theta}.
    \label{eq:HS-uh-outside}
\end{align}
Finally, the total velocity is the sum of electro-osmotic and hydraulic terms:
\begin{equation}
    \mathbf{u}_{1,2} = \mathbf{u}^{1,2}_h + \mathbf{u}^{1,2}_{eo}.
    \label{eq:utotal}
\end{equation}

The velocity field in equation \eqref{eq:utotal} is the sum of a uniform background flow and a dipole-like term due to nonuniformity in the gap thickness. This velocity field depends on two nondimensional parameters: the ratio of the gap thickness between inside and outside $h = H_1/H_2$, and the ``electro-osmotic coupling coefficient'':
\begin{equation}
    \alpha_{eo} = \frac{U_{eo}}{U},
    \label{eq:HS-alpha}
\end{equation}
which measures the relative strength of electro-osmotic velocity with respect to the total velocity. For a fixed geometry, the electro-osmotic coupling coefficient can be tuned independently and affects the flow field. When $\alpha_{eo} > 0$, the electro-osmotic and far-field velocities are in the same direction, whereas negative values ($\alpha_{eo} < 0$) indicate opposite directions. Irrespective of the sign of $\alpha_{eo}$, flow reversal is expected when the electro-osmotic velocity is sufficiently strong, as illustrated in figure \ref{fig:hele-shaw}. To characterize the flow behavior, we consider the $\theta$-component of the total velocity along $\theta = \pi/2$:
\begin{align}
    \frac{u_1(r)}{U} &= 2\left(\frac{\alpha_{eo}}{1+h} + \frac{(1-\alpha_{eo})h^2}{1+h^3}\right), \quad r < a,
    \label{eq:HS-u1} \\
    \frac{u_2(r)}{U} &= 1 + \frac{a^2}{r^2} \left(\alpha_{eo}\frac{1-h}{1+h} + (1-\alpha_{eo})\frac{1-h^3}{1+h^3}\right), \quad r>a.
    \label{eq:HS-u2}
\end{align}
We detect the presence of circulation if $u_1(a)u_2(a) < 0$, \ie the $\theta$-component of velocity switches sign across the disk region. Figure \ref{fig:hele-shaw} illustrates the regions in the $(h, \alpha_{eo})$ phase space for which flow circulation is possible:
\begin{equation}
    \textbf{circulation possible:}\quad \alpha_{eo} \le \frac{-h^2}{1-h}\, \quad \text{or} \quad \alpha_{eo} \ge \frac{1}{h (1-h)}.
\end{equation}
Note that when $\alpha_{eo} > 0$, the direction of the dipolar flow field is aligned with the pressure-driven flow and we say the circulation is ``positive''. Conversely, when $\alpha_{eo} < 0$, the dipolar and pressure-driven flow fields are in the opposite direction and we call the circulation to be ``negative''. The extent of the circulation region is computed by requiring that only the radial (when $\alpha_{eo} < 0$) or both components (when $\alpha_{eo} > 0$) of velocity are zero at $(r,\pi/2)$:
\begin{align}
    \frac{r}{a} &= \sqrt{\frac{1-h^3}{1+h^3} - 2\alpha_{eo}\frac{h(1-h)}{1+h^3}}, \quad \alpha_{eo} < \frac{-h^2}{1-h},
    \label{eq:HS-alpha-neg} \\
    \frac{r}{a} &= \sqrt{2\alpha_{eo}\frac{h(1-h)}{1+h^3} - \frac{1-h^3}{1+h^3}}, \quad \alpha_{eo} > \frac{1}{h (1-h)}.
    \label{eq:HS-alpha-pos}
\end{align}

Here, we have shown that electro-osmotic flows can generate circulating regions in a Hele-Shaw cell with a nonuniform gap thickness. Although the analysis was more involved than the previous section, the fundamental physical picture remains unchanged: electro-osmotic flows create adverse pressure gradient, which subsequently drive hydraulic flows in the opposite direction. This can lead to flow reversal and circulation in regions where the gap thickness is nonuniform. The spatial extent of this circulating region scales linearly with the size of nonuniformity.

\section{Electrokinetic Transport in Random Porous Media}  \label{sec:porous-media}
\subsection{Electrokinetic Transport in Porous Media}
The fully coupled electrokinetic transport equations describe a linear relationship between fluxes, \ie fluid velocity, $\mathbf{u}$, and electric current density, $\mathbf{i}$, and thermodynamic driving forces, \ie pressure, $p$, and electric potential, $\phi$. Combined with statements of mass and charge conservation, the governing equations read:
\begin{align}
    \divg{\mathbf{F}} = \mathbf{0}, & \quad \mathbf{F} = -\mathbb{K}\nabla \mathbf{\Phi},
    \label{eq:coupled_ek_porous}
\end{align}
where we use $\mathbf{F} = (\mathbf{u}, \mathbf{i})^\mathsf{T}$ and $\mathbf{\Phi} = (p, \phi)^\mathsf{T}$ notation for brevity and $\mathbb{K}$ is the electrokinetic coupling tensor:
\begin{equation}
    \mathbb{K} = \left(\begin{array}{cc}
        K_h    & K_{eo} \\
        K_{eo} & K_e
    \end{array} \right).
    \label{eq:K_tensor}
\end{equation}
Equation \eqref{eq:coupled_ek_porous} indicates that the fluid velocity is the sum of hydraulic, $\mathbf{u}_h = -K_h \nabla p$, and electro-osmotic, $\mathbf{u}_{eo} = -K_{eo} \nabla \phi$, terms. Similarly, the electrical current is the sum of Ohmic, $\mathbf{i}_e = -K_e \nabla \phi$, and streaming current, $\mathbf{i}_{sc} = -K_{eo} \nabla p$, terms. The symmetry of the electrokinetic tensor, $\mathbb{K}$, is a result of microscopic reversibility and can be directly verified using Stokes and Poisson-Nernst-Planck equations. Here, $K_h(\mathbf{x}) = k(\mathbf{x})/\mu$ is the hydraulic conductivity where $k(\mathbf{x})$ is the permeability field. Since we focus on thin EDL approximation, the electrokinetic mobility, $K_{eo}$, is given via the Helmholtz-Smoluchowski relation, $K_{eo} = -\varepsilon \zeta/\mu$, and the electrical conductivity, $K_e$, is equal to that of electrolyte conductivity, \ie $K_e = \sigma$. Note that other than the permeability field, the remaining coefficients are assumed to be uniform in the domain.

The strength of electrokinetic coupling is measured in terms of the nondimensional coupling coefficient:
\begin{equation}
    \alpha = \frac{K_{eo}^2}{K_h K_e},
    \label{eq:alpha}
\end{equation}
which also controls the efficiency of electrokinetic energy conversion \cite{van2006electrokinetic}.
The second law of thermodynamic requires that $0 \le \alpha < 1$ ( $\det\mathbb{K}\ge0$). In the limit of thin EDLs, the coupling is typically weak, \ie $\alpha \ll 1$. In this ``weak coupling'' limit, only the streaming current or electro-osmotic flow can be large but not both at the same. This can be verified by introducing two nondimensional parameters:
\begin{equation}
    \alpha_{eo} = \frac{K_{eo} I}{K_e U} \sim U_{eo}/U, \quad \alpha_{sc} = \frac{K_{eo} U}{K_h I} \sim I_{sc}/I,
    \label{eq:alphas}
\end{equation}
which measure the relative importance of electro-osmotic velocity and streaming current, respectively. By comparing to equation \eqref{eq:alpha}, it is clear that $\alpha = \alpha_{eo} \alpha_{sc}$ and therefore, in the limit of weak coupling, both effects cannot be strong at the same time.

In our numerical simulations, we solve the fully coupled equations \eqref{eq:coupled_ek_porous}. However, to better understand the role of electro-osmotic flow, we can simplify these equations in the limit of weak streaming current, \ie $\alpha_{sc} \ll 1$:
\begin{align}
    \divgp{K_h(\mathbf{x}) \nabla p} + K_{eo} \nabla^2 \phi &=0, \label{eq:simple-pressure} \\
    K_e \nabla^2 \phi &=0, \label{eq:simple-potential}
\end{align}
Equation \eqref{eq:simple-pressure} can further be simplified by using equation \eqref{eq:simple-potential}:
\begin{align}
    \divgp{\frac{k(\mathbf{x})}{\mu}\nabla p} &= 0, \label{eq:simplified_pressure} \\
    \nabla^2\phi &= 0. \label{eq:simplified_potential}
\end{align}
The apparent decoupling of pressure and potential fields is again due to assuming uniform electrokinetic and electric conductivity coefficients. In a more realistic scenario, all of the terms in the conductivity tensor in equation \eqref{eq:K_tensor} could be heterogeneous and the two fields remained fully coupled. Variation in the zeta potential is possible due to heterogeneous surface chemistry. Furthermore, spatio-temporal variations in the electrolyte properties, such as pH and salt concentration, can lead to nonlinear electrokinetic response and formation of deionization shocks, which cause heterogeneous zeta potential and solution conductivity \cite{alizadeh2019impact}. Nevertheless the pressure and electric potential are still coupled through the boundary conditions when fluxes are prescribed:
\begin{align}
    -\frac{k(\mathbf{x})}{\mu} \hat{\mathbf{n}} \cdot \nabla p &= U_n + K_{eo} \hat{\mathbf{n}} \cdot \nabla \phi,
    \label{eq:pressure_bc} \\
    -K_e \hat{\mathbf{n}} \cdot \nabla \phi &= I_n, \label{eq:potential_bc}
\end{align}
where $\hat{\mathbf{n}}$ denotes the normal to the boundary, and $U_n$ and $I_n$ are the total velocity and electric current density, respectively. Combining equations \eqref{eq:pressure_bc} and \eqref{eq:potential_bc} yields the following simple boundary condition for the pressure:
\begin{equation}
    -\frac{k(\mathbf{x})}{\mu} \hat{\mathbf{n}} \cdot \nabla p = U_n - \frac{K_{eo}}{K_e} I_n = U_n (1 - \alpha_{eo}).
    \label{eq:pressure_bc_simple}
\end{equation}

Equation \eqref{eq:pressure_bc_simple} suggests that the pressure field and fluid velocity can be directly controlled by adjusting the current density at the boundaries. Finally, once the pressure and electric potential are known, the fluid velocity in the domain is given via:
\begin{equation}
    \mathbf{u} = -\frac{k(\mathbf{x})}{\mu} \nabla p - K_{eo} \nabla \phi
               = (1-\alpha_{eo})\mathbf{u}_0 + \mathbf{u}_{eo},
    \label{eq:domain-velocity}
\end{equation}
where $\mathbf{u}_0$ is the fluid velocity for the same problem but at zero electric field and $\mathbf{u}_{eo} = -K_{eo} \nabla \phi$. Because $\alpha_{eo}$ is controlled by the electric current, it can be tuned independent of the fluid velocity. Remarkably, when $\alpha_{eo} = 1$, the fluid velocity is independent of pressure field and is entirely dictated by the electro-osmotic velocity.

\subsection{Random Field Generation}
\begin{figure}
    \centering
    \includegraphics[width=\columnwidth]{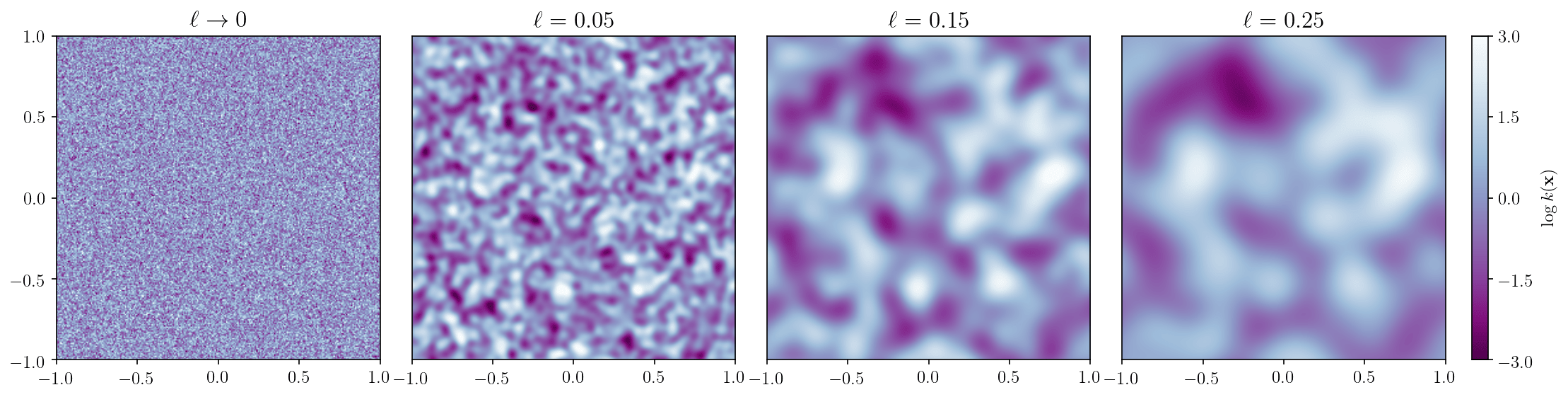}
    \caption{A sequence of random fields generated for a Gaussian autocorrelation function with $\xi_0 = 1$ and different values of correlation length. When $\ell \rightarrow 0$, the random field appears as a white noise but increasing the correlation length reduces large variations in nearby points and results in a smoothly varying random field.}
    \label{fig:random-fields}
\end{figure}
We use random fields to represent heterogeneous properties in a random porous medium. Specifically, we assume the permeability field is given via:
\begin{equation}
    k(\mathbf{x}) = k_0 \exp(z(\mathbf{x})),
    \label{eq:kdef}
\end{equation}
where $k_0$ is a reference value and $z(\mathbf{x})$ is a random field with a known autocorrelation function $\xi(\mathbf{x})$:
\begin{equation}
    \xi(\mathbf{x}) = \int z(\mathbf{x'}) z(\mathbf{x}' - \mathbf{x}) \, \mathrm{d \mathbf{x}'},
    \label{eq:autocorr}
\end{equation}
We further assume the random field $z(\mathbf{x})$ to be statistically isotropic, \ie $\xi(\mathbf{x}) = \xi(|\mathbf{x}|$), although this restriction can be easily lifted. Different correlation functions may be assumed depending on the statistical properties of the medium. Here, we assume Gaussian correlations, \ie
\begin{equation}
    \xi(|\mathbf{x}|) = \xi_0 \exp (-|\mathbf{x}|^2/\ell^2),
    \label{eq:gaussian_corr}
\end{equation}
where $\xi_0$ controls the variance of the random field $z(\mathbf{x})$. Indeed, we have $\xi_0 = \sigma^2 + \mu^2$ where $\sigma^2$ and $\mu$ are the variance and mean of the random field $z(\mathbf{x})$. The ``correlation length'' $\ell$, defines a length-scale over which the correlations in the random field decay and could be understood as defining a ``feature size'' in the random field (see figure \ref{fig:random-fields}).

We use Fast Fourier Transform (FFT) to generate random fields efficiently. This approach is similar to the algorithm presented by authors in \cite{shinozuka1972digital, shinozuka1991simulation} and based on the Wiener-Khinchin theorem, which states that the autocorrelation and power spectral density functions are Fourier transform pairs:
\begin{equation}
    \xi(\mathbf{x}) \xleftrightarrow{\mathcal{F}} S(\mathbf{k}):\, \mathcal{F} \xi(\mathbf{x}) = |\mathcal{F} z(\mathbf{x})|^2 = S(\mathbf{k}),
    \label{eq:wiener-khinchin}
\end{equation}
where $\mathcal{F}$ denotes the Fourier transform and $\mathbf{k}$ is the wave vector. For the Gaussian autocorrelation function defined in \eqref{eq:gaussian_corr}, the power spectral density is given via:
\begin{equation}
    S(|\mathbf{k}|) = \int \xi(|\mathbf{x}|) \,e^{-i\mathbf{k}\cdot\mathbf{x}} \,\mathrm{d}\mathbf{x} = \pi \xi_0 \ell^2 \exp\left(-\ell^2 |\mathbf{k}|^2/4\right).
    \label{eq:psd}
\end{equation}
With a known power spectral density, we generate the random field $z(\mathbf{x})$ by taking inverse Fourier transform while randomizing the phase of individual modes:
\begin{equation}
    z(\mathbf{x}) = \mathcal{F}^{-1}\left[\sqrt{S(|\mathbf{k}|)}\, e^{i\theta(\mathbf{k})}\right] = \frac{1}{4\pi^2}\int \sqrt{S(|\mathbf{k}|)}\,e^{i\theta(\mathbf{k})} \,e^{i\mathbf{k}\cdot\mathbf{x}} \,\mathrm{d}\mathbf{k},
    \label{eq:inverse-fourier}
\end{equation}
where $\theta(\mathbf{k})$ is the random phase angle for mode $\mathbf{k}$, which is drawn from a uniform distribution, \ie $\theta(\mathbf{k}) \sim \mathcal{U}(0, 2\pi)$. We also require that $\theta(-\mathbf{k}) = \theta(\mathbf{k})$, so that $z(\mathbf{x})$ is real valued. Equation \eqref{eq:inverse-fourier} could be understood as a superposition of plane waves with amplitude $\sqrt{S(|\mathbf{k}|)}$ and randomized phase angles. Alternatively, equation \eqref{eq:inverse-fourier} could also be understood as smoothing a white noise field using a filter whose power density is $S(|\mathbf{k}|)$. Figure \ref{fig:random-fields} illustrates a sequence of random fields generated using equation \eqref{eq:inverse-fourier} with $\xi_0=1$ and different correlation lengths.

\subsection{Numerical Simulations}
\begin{figure}[t!]
    \centering
    \includegraphics[width=\columnwidth]{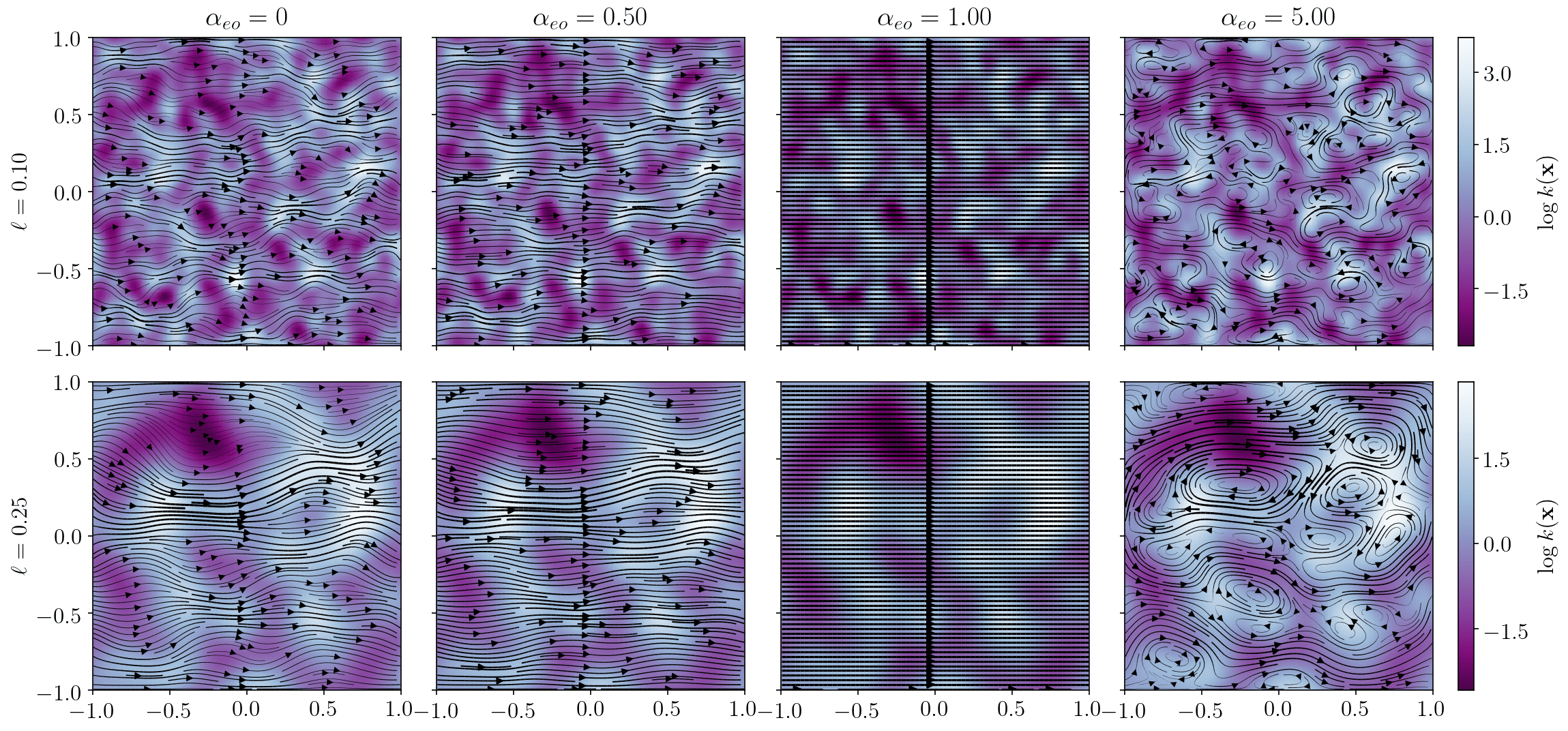}
    \caption{Effects of electro-osmotic coupling coefficient ($\alpha_{eo}$) and correlation length ($\ell$) on the vortex size. The solid lines are the streamlines and the background color is logarithm of the permeability field, \ie $z(\mathbf{x}) = \log k(\mathbf{x})$. When $\alpha_{eo} = 1$, the flow field is entirely determined by the electro-osmotic flow, which is uniform due to uniform electric field (\cf equation \eqref{eq:domain-velocity}). }
    \label{fig:constant-velocity-forward}
\end{figure}
We numerically solve the system of equations \eqref{eq:coupled_ek_porous} in nondimensional form. This is achieved by introducing the following nondimensional variables
\begin{equation}
    \mathbf{\hat{x}} = \frac{\mathbf{x}}{L}, \quad
    \hat{k}(\mathbf{x}) = \frac{k(\mathbf{x})}{k_0}, \quad
    \hat{p} = \frac{p k_0}{U L \mu}, \quad
    \hat{\phi} = \frac{\phi K_{e}}{I L}, \quad
    \mathbf{\hat{u}} = \frac{\mathbf{u}}{U}, \quad
    \mathbf{\hat{i}} = \frac{\mathbf{i}}{I}, \quad
    \label{eq:ref_values}
\end{equation}
where $L$ is a reference macroscopic size, $k_0$ is a reference permeability coefficient, and $U$ and $I$ are reference velocity and electric current density values. For brevity, we will omit the hat notation in the remaining of this article and treat all variables as nondimensional unless otherwise noted. The resulting nondimensional equations are given via:
\begin{align}
    \divgp{k(\mathbf{x}) \nabla p} + \alpha_{eo} \nabla^2 \phi &= 0, \label{eq:nondim-pressure} \\
    \alpha_{sc} \nabla^2 p + \nabla^2 \phi &= 0, \label{eq:nondim-potential}
\end{align}
with the coupling coefficients defined in equation \eqref{eq:alphas}.

As discussed earlier, the nondimensional coupling coefficient,
\begin{equation}
    \alpha = \alpha_{eo} \alpha_{sc} = \frac{K_{eo}^2}{K_e K_h} = \frac{\varepsilon^2\zeta^2 k_0}{\mu \sigma}, \label{eq:alpha-detailed}
\end{equation}
measures the overall coupling between pressure and potential field and is often very small in practice. For instance, assuming a reference permeability $k_0 = 10 \,\mathrm{mD} \approx 10^{-14}\, \mathrm{m^2}$, $\zeta = -50 \,\mathrm{mV}$, and $1\, \mathrm{mM}$ monovalent binary aqueous electrolyte with $\sigma \approx 7.5\,\mathrm{mS\,m^{-1}}$ and $\varepsilon \approx 7\times10^{-10}\,\mathrm{F\,m^{-1}}$, the coupling coefficient is $\alpha \approx 1.6 \times 10^{-2}$.

As discussed in the previous section, we expect the vortical structures to first appear when $|\alpha_{eo}| \sim 1$. From equation \eqref{eq:alphas}, this condition corresponds to a critical injection ratio of
\begin{equation}
    \left(I/U\right)_{cr} = \frac{K_e}{K_{eo}} \approx 212 \frac{\mathrm{mA\,cm^{-2}}}{\mathrm{cm\,s^{-1}}}.
    \label{eq:Icr}
\end{equation}
For a velocity of $U \approx 0.1 \, \mathrm{mm\,s^{-1}}$, the critical current is relatively small $I_{cr} \approx 2.1 \,\mathrm{mA\,cm^{-2}}$. We also note that the right hand side of equation \eqref{eq:Icr} is reciprocal of an important quantity in the geophysics community, \ie the ``electrokinetic coupling coefficient'', $C = K_{eo}/K_e$\cite{sill1983self}. This quantity is often determined directly through streaming potential measurements. For many reservoirs, measured values fall in the range of $C \sim 10^{-9} - 10^{-5}\, \mathrm{V\,Pa^{-1}}$ depending on the pore fluid salinity, with typical value of $C_v \approx 10^{-6}$ at salinity level of $c_f\approx 1\,\mathrm{mM}$ \cite{glover2012streaming}.

We solve the coupled equations \eqref{eq:nondim-pressure} and \eqref{eq:nondim-potential} in a square domain $\Omega = [-1, 1]^2$ subjected to prescribed flux values at the left and right boundary,
\begin{align}
    -k(\pm 1, y) \left.\frac{\partial p}{\partial y}\right|_{x=\pm 1} - \alpha_{eo} \left.\frac{\partial \phi}{\partial y}\right|_{x=\pm 1} &= 1,
    \label{eq:nondim-pressure-bc} \\
    -\alpha_{sc} \left.\frac{\partial p}{\partial y}\right|_{x=\pm 1} - \left.\frac{\partial \phi}{\partial y}\right|_{x=\pm 1} &= 1,
    \label{eq:nondim-potential-bc}
\end{align}
and no flux conditions at the top and bottom boundaries. Figure \ref{fig:constant-velocity-forward} illustrates the formation of vortical structures for different values of electro-osmotic coupling coefficient and correlation length. At zero electro-osmotic coupling, the fluid flow is dictated by the pressure field and the heterogeneous permeability field. As the coupling coefficient is increased, the velocity field initially becomes uniform, assisted by the electro-osmotic flow in the same direction. From equation \eqref{eq:domain-velocity}, when $\alpha_{eo} = 1$, the flow field is entirely determined by the electro-osmotic component, which is uniform due to uniform electric field. When $\alpha_{eo} > 1$, strong electro-osmotic flows through regions of low permeability create backward pressure-driven flow through regions of high permeability, resulting in a circulating flow field. The size of vortical structures is therefore directly related to the separation distance between low and high permeability values, which is controlled by the correlation length parameter $\ell$.

\begin{figure}[t!]
    \centering
    \includegraphics[width=\columnwidth]{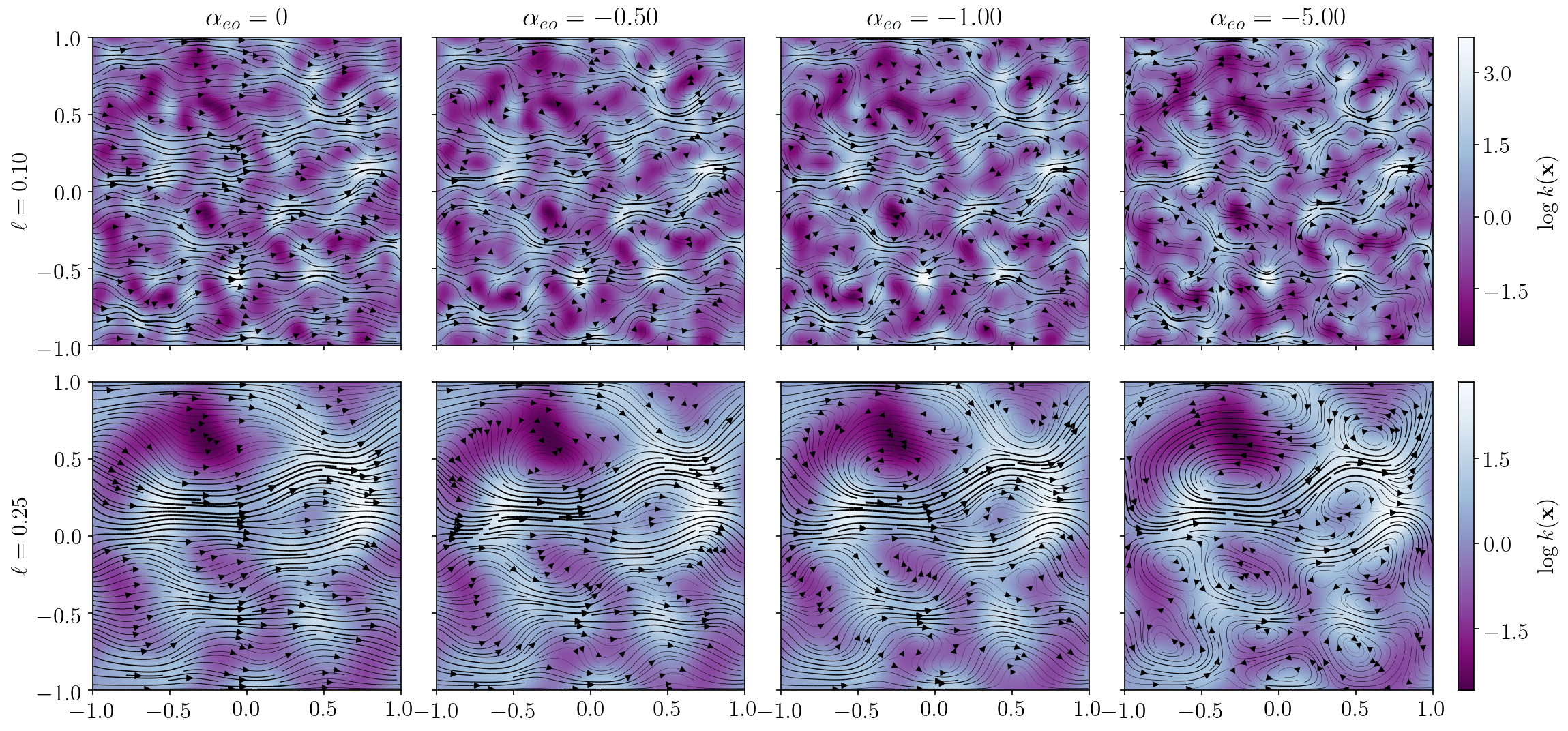}
    \caption{Vortical structure persist when the electro-osmotic and pressure driven flows are in the opposite directions. The solid lines are the streamlines and the background color is logarithm of the permeability field, \ie $z(\mathbf{x}) = \log k(\mathbf{x})$. Note that, unlike figure \ref{fig:constant-velocity-forward}, the vortical structures appear even when $|\alpha_{eo}| < 1$.}
    \label{fig:constant-velocity-reverse}
\end{figure}

Figure \ref{fig:constant-velocity-reverse} illustrates the flow patterns when the electro-osmotic and pressure driven flows are in the opposite direction. This is simulated by reversing the sign of the electro-osmotic coupling coefficient, which has the same effect as reversing the sign of electric current in boundary condition \eqref{eq:nondim-potential-bc}. Unlike figure \ref{fig:constant-velocity-forward}, vortical structures can appear for any value of electro-osmotic coupling, even when $|\alpha_{eo}| < 1$. Nevertheless, the number of vortices grows as the magnitude of electro-osmotic coupling is increased. These results indicate that vortex generation is a consequence of nonuniform competition between electro-osmotic and pressure driven flows and occurs when electro-osmotic flows are sufficiently strong.

\begin{figure}
    \centering
    \includegraphics[width=0.85\columnwidth]{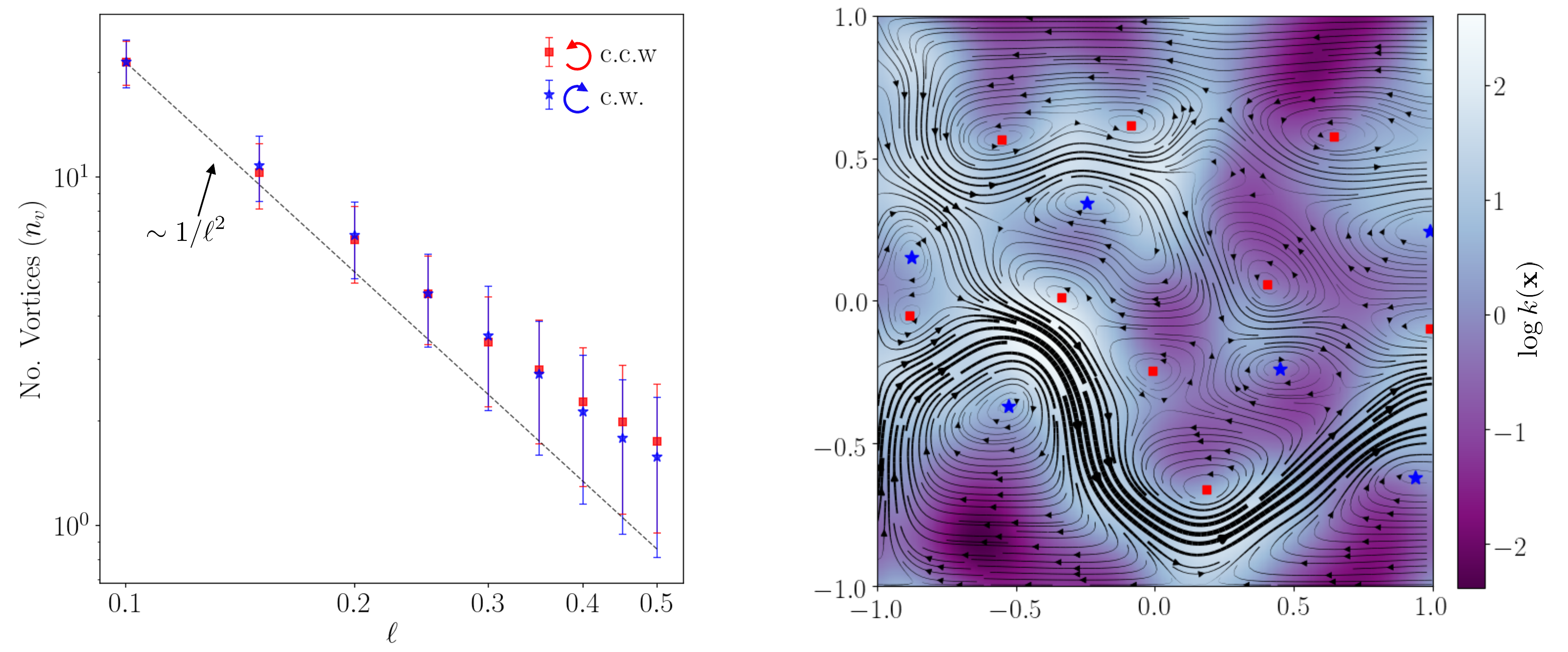}
    \caption{Statistical analysis of vortex size. \textbf{Left:} The number of vortices (both clockwise and counter-clockwise) increases as the correlation length is decreased and scales as $n_v \sim \ell^{-2}$ for sufficiently small correlation lengths. The electro-osmotic coupling coefficient was set to $\alpha_{eo} = -3$. The symbols represent ensemble averages and error bars are one standard deviation involving 100 simulations for each value of correlation length. The discrepancy at larger correlation length might be due to finite size effects and strong interaction of vortices in the periodic domain. \textbf{Right:} A representative example illustrating the applicability of our algorithm in detecting vortex centers for $\ell = 0.2$.}
    \label{fig:statistical-analysis}
\end{figure}

From figures \ref{fig:constant-velocity-forward} and \ref{fig:constant-velocity-reverse}, it appears that the vortex size directly scales with the correlation size. To quantify this relationship, we generate different realizations for fixed correlation length and compute the ensemble average number of vortices. We define a vortex center as the maximum or minimum of the stream function, $\psi$:
\begin{equation}
    \mathbf{u} = \left(\frac{\partial \psi}{\partial y}, -\frac{\partial \psi}{\partial x}\right)^\mathsf{T}.
\end{equation}
In this definition, clockwise vortices are associated with the minima of the stream function (negative vorticity) and counter-clockwise vortices are associated with the maxima of the stream function (positive vorticity). To detect vortices, first a collection of candidate critical points are selected by thresholding the velocity magnitude, \ie $|\mathbf{u}| < 5\times10^{-3}\, U_{\max}$ where $U_{\max}$ is the maximum velocity magnitude in the domain. Next, the saddle points are rejected by computing the eigen-values of the Hessian matrix:
\begin{equation}
    \mathsf{H} = \left(\begin{array}{cc}
        \frac{\partial^2\psi}{\partial x^2} & \frac{\partial^2\psi}{\partial x \partial y} \\
        \frac{\partial^2\psi}{\partial x \partial y} & \frac{\partial^2\psi}{\partial y^2}
    \end{array}\right),
\end{equation}
given via:
\begin{equation}
    \lambda^2 - \mathrm{tr}(\mathsf{H}) \lambda + \det(\mathsf{H}) = 0.
\end{equation}
The critical points of the stream function are then classified accordingly:
\begin{equation}
    \mathbf{maximum:}\: \lambda_1 > 0, \lambda_2 > 0, \quad
    \mathbf{minimum:}\: \lambda_1 < 0, \lambda_2 < 0, \quad
    \mathbf{saddle:}\: \lambda_1 \lambda_2 < 0.
\end{equation}
Figure \ref{fig:statistical-analysis} illustrates the ensemble averaged number of vortices as a function of correlation length. The number of vortices ($n_v$) increases as the correlation size is decreased, and for sufficiently small values, scales as $n_v \sim \ell^{-2}$. This scaling further suggests that the average vortex size scales with the correlation length, \ie $\ell_v \sim 1/\sqrt{n_v} \sim \ell$. Interestingly, the appearance of adjacent counter-rotating vortices resembles the vortical patterns often observed in ``bacterial turbulence'' \cite{dunkel2013fluid}. Similar alternative charge ordering is also observed in the ionic positions of cations and anions in ionic liquids \cite{levy2019spin}. It would be interesting to see if similar ``antiferromagnetic'' ordering also applies for electrokinetic vortices and whether the ideas presented in \cite{levy2019spin} can be used to reconstruct the flow field.

\subsection{Electrokinetic Mixing}
\begin{figure}[t!]
    \centering
    \includegraphics[width=\columnwidth]{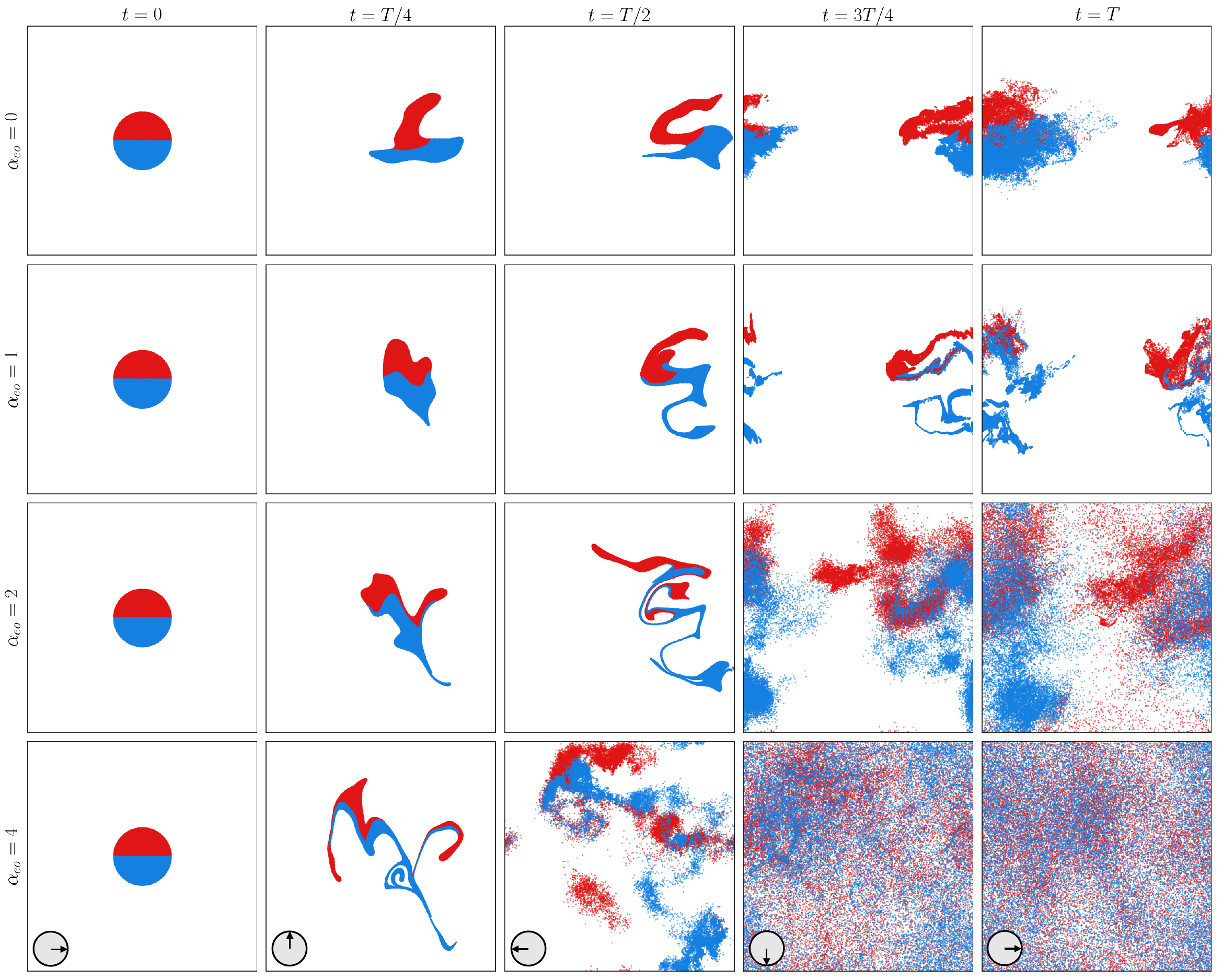}
    \caption{Effect of electro-osmotic vortices on mixing. A rotating electric field, shown by the small arrow in the last row, generates an unsteady flow field that causes strong mixing. This is demonstrated by advection of $50,000$ Lagrangian markers with the local flow field. Mixing is enhanced as the strength of electro-osmotic flow is increased. The background permeability field (omitted for clarity) is the same in all cases and corresponds to a random field with $\ell = 0.15$.}
    \label{fig:mixing}
\end{figure}
We finish this paper by pointing that the ideas presented in this article might be helpful in controlling and enhancing mixing in porous media. Mixing in porous media is important in engineering applications and geological processes such as secondary oil recovery, contaminant transport, and soil and groundwater remediation \cite{steefel2005reactive, dentz2011mixing}. When reactions are fast, transport processes are diffusion-limited and will benefit from enhanced mixing. Recently, there has been renewed interest in the role of heterogeneity in enhancing the mixing rate \cite{nicolaides2015impact}. In a heterogeneous media, mixing is improved due to increased velocity fluctuation and material stretching \cite{le2013stretching, amooie2017mixing}. Nevertheless, in passive mixing, it is difficult to control the degree of mixing due to small Reynolds number.

As we have shown in this paper, electrokinetic phenomena can generate strong vortical flows, which are easy to control. Combined with a suitable control strategy, electrokinetic vortices could enhance chaotic mixing in porous media and improve existing remediation techniques \cite{shapiro1993removal,probstein1993removal}. Note that our proposed idea is different from chaotic electrokinetic flows in micro-fluidic channels \cite{posner2006convective, posner2012electric} and near ion-selective membranes \cite{rubinstein2000electro, zaltzman2007electro, druzgalski2013direct}, which are generated due to nonlinear interactions.  It is also different from stochastic electrotransport in porous media subjected to randomly fluctuating electric fields \cite{kim2015stochastic}, which are used to deliver and purge chemicals in preparation for imaging of large tissue sample, such as whole mouse brains \cite{murray2015simple,ku2016multiplexed,mano2018whole}, although the dispersion caused by any electrokinetic vortex patterns could ideally be added to effective diffusion coefficient associated with random electrophoretic drift.

To provide a concrete example, consider a rotating electric field defined as:
\begin{equation}
    \mathbf{E} = E_0 (\cos(\omega t), \sin(\omega t))^\mathsf{T},
\end{equation}
with constant frequency $\omega = 2 \pi/T$. For simplicity, we assume the period is equal to the advection time-scale, \ie $T = L/U$. This electric field can be realized by placing two sets of perpendicular electrodes, with AC signals that have a phase shift of $\pi/2$. The rotating electric field generates a periodic unsteady flow field, which can cause strong mixing, especially in applications where the field rotation is stochastic, caused by random motion of a porous sample \cite{kim2015stochastic}. We characterize the effectiveness of electro-osmotic mixing via Lagrangian advection of tracers with the local velocity field:
\begin{equation}
    \frac{\mathrm{d} \mathbf{x}}{\mathrm{d} t} = \mathbf{u}(\mathbf{x}, t).
\end{equation}
Figure \ref{fig:mixing} illustrates consecutive snapshots of advecting $N=50,000$ tracers during one period of rotating field for different values of electro-osmotic coupling coefficient (see supplementary information for the movie). For all simulations, the correlation length of the random permeability field was set to $\ell = 0.15$. The tracers are initially distributed uniformly and randomly inside a disk region and are colored to better demonstrate the mixing process. As the strength of electro-osmotic flow is increased, the mixing is enhanced due to stronger vortical flow. Although mixing still occurs without electric field (top row) due to spatial velocity fluctuations, electro-osmotic vortices significantly enhance the mixing as evident in the bottom row.

\section{Conclusions}
In this article, we have demonstrated that electro-osmotic flows in heterogeneous porous media can lead to the formation of vortex patterns. This phenomenon occurs due to competition between pressure-driven and electro-osmotic flows. Because the pressure-driven and electro-osmotic flows scale differently with the pore size, large internal pressure gradients are created when electro-osmotic flow is pushed through tight pores. When the permeability field is nonuniform, the generated pressure creates backward flow through regions of high permeability. Although this mechanism has been widely known for a single pore, here we demonstrated that a similar flow reversal also occurs on the macroscopic scale, much larger than individual pores. Through detailed analyses, we showed that the spatial extent of vortical structures directly scales with the length scale of the heterogeneity in the permeability field. We also introduce a nondimensional parameter, the electro-osmotic coupling coefficient, $\alpha_{eo}$, which describes the relative strength of electro-osmotic flow and controls the flow pattern. When $\alpha_{eo} \sim 1$, the electro-osmotic flow is comparable to pressure-driven flow and vortical structures first appear. Importantly, the electro-osmotic coupling coefficient may be tuned by adjusting the applied current or electric field, independently from the flow rate and pressure difference.

The analysis we have presented here may be extended in several ways. First, we have assumed that only the permeability field is heterogeneous while other material properties are uniform in the domain. In practice, $\zeta$-potential could also be heterogeneous due to nonuniform chemical composition. Moreover, several material properties could change dynamically with flow field. In single-phase flows, variation in salt concentration or pH can lead to changes in electrolyte conductivity and $\zeta$-potential, leading to nonlinear coupling and formation of deionization shock waves. The problem is even more complex in two-phase flows, where the immiscibility can lead to unequal partitioning of ions and nonuniform electrokinetic response. Recent works in Hele-Shaw cells suggest these interactions can be exploited to control viscous fingering, but their applicability to porous media is still unknown. Both the nonlinear response and two-phase electrokinetic flow in porous media are novel problems that require further theoretical and experimental investigations.

Finally, we have shown that strong electro-osmotic flows generate vortical structures that promote fluid mixing. Although several mixing strategies have been proposed in micro-fluidic systems, controllable mixing in porous media is more difficult to achieve. This is because such techniques often rely on geometrical modifications, which are usually not possible in a porous medium. By contrast, electrokinetic mixing relies on the inherent heterogeneity of the medium and is easily controlled externally. The ideas presented in this article may therefore be exploited in applications where controllable mixing is desired, \eg in enhancing reaction rates in porous electrodes, improving the effective diffusivity of molecules in biological tissues, and diluting contaminants in soil during electrokinetic remediation.

\section*{Acknowledgments}
This work was supported by a fund from Aramco Americas Company under contract number A-0434-2017.
\bibliography{refs}

\end{document}